\documentclass[traditabstract]{aa} 
\usepackage{graphicx}
\usepackage{longtable}
\usepackage{amsfonts}
\usepackage{rotating}
\RequirePackage{amssymb}
\RequirePackage{latexsym}

\usepackage{natbib}

\def\umy{\hbox{\it u--y\/~}}

\def\bmy{\hbox{\it b--y\/~}}
\def\feh{\hbox{\rm [Fe/H]~}}
\def\mh{\hbox{\rm [M/H]~}}
\def\zx{\hbox{\rm Z/X}}

\def\afe{\hbox{\rm [$\alpha$/Fe]}}
\renewcommand{\deg}{\mbox{$^{\circ}$}}

\newcommand{\strom}{\mbox{Str\"omgren~}}

\usepackage{txfonts}
\begin{document}
   \title{A new visual -- near-infrared diagnostic to estimate the metallicity of cluster and field dwarf stars}


   \author{A. Calamida\inst{1}
   \and M. Monelli\inst{2,3}
   \and A. P. Milone\inst{2,3}
  \and G. Bono\inst{4}
  \and A. Pietrinferni\inst{5}
  \and E. P. Lagioia\inst{4}}

   \institute{INAF-Osservatorio Astronomico di Roma, 
              Via Frascati 33, 00040, Monte Porzio Catone, Italy \\
              \email{annalisa.calamida@oa-roma.inaf.it}
\and IAC - Instituto de Astrofisica de Canarias, Calle Via Lactea,
E38200 La Laguna, Tenerife, Spain \\
\email{monelli@iac.es; milone@iac.es}
\and Department of Astrophysics, University of La Laguna, E-38200 La Laguna, Tenerife, Spain
\and Universit\`a di Roma Tor Vergata, Via della Ricerca Scientifica 1,
00133 Rome, Italy \\
\email{Giuseppe.Bono@roma2.infn.it; eplagioia@roma2.infn.it}
\and INAF-Osservatorio Astronomico di Collurania, via M. Maggini, 
64100 Teramo, Italy \email{adriano@oa-teramo.inaf.it} }

   \date{}

 
  \abstract
   {We present a theoretical calibration of a new metallicity diagnostic based 
    on the \strom index $m_1$ and on visual -- near-infrared (NIR) colors to estimate 
    the global metal abundance of cluster and field dwarf stars.
    To perform the metallicity calibration we adopt $\alpha$-enhanced 
    evolutionary models transformed into the observational plane by using 
    atmosphere models computed adopting the same chemical mixture.
    We apply the new visual - NIR Metallicity--Index--Color (MIC) relations to two different 
    samples of field dwarfs and we find that the difference between photometric 
    estimates and spectroscopic measurements is on average smaller than 
    0.1 dex, with a dispersion smaller than $\sigma$ = 0.3 dex. 
    We apply the same MIC relations to a metal-poor (M~92) and a metal-rich 
    (47~Tuc) globular cluster. We find a peak of -2.01$\pm$0.08 ($\sigma$ = 0.30 dex) 
    and -0.47$\pm$0.01 ($\sigma$ = 0.42 dex), respectively.}


   \maketitle
%

\section{Introduction}

The intermediate-band \strom photometric system \citep{strom66}
has, for stars with spectral types from A to G, several indisputable 
advantages when compared with broad-band photometric systems.

{\em i\/}) The ability to provide robust estimates of intrinsic stellar parameters 
such as the metal abundance (the $m_1=(v-b)-(\bmy)$ index, 
\citealt[hereafter CA07]{twa00, hilker00,io07}), the surface gravity (the $c_1=(u-v)-(v-b)$ index), 
and the effective temperature 
(the $H_{\beta}$ index, \citealt{nissen88, ols88, twa00}). 
The $H_{\beta}$ index is marginally affected by reddening, and therefore 
can also be compared to a simple color such as \bmy\  to provide individual estimates 
of reddening corrections \citep{nissen91}. The same outcome applies to 
the reddening free $[c_1]$ index, and indeed theoretical and empirical 
evidence \citep{ste91, nissen94, io05} suggests that 
a color such as \umy\ compared to $[c_1]$ --which is a temperature index
for stars hotter than 8,500 K-- provides a robust reddening index for blue 
horizontal branch stars.

{\em ii\/}) The use of the $m_1/c_1$ versus color plane can also be safely 
adopted to distinguish cluster and field stars \citep{twa00, rey04, faria, 
aden09, arnadottir}.    

{\em iii\/}) Accurate \strom photometry can also be adopted 
to constrain the ensemble properties of stellar populations in complex 
stellar systems like the Galactic bulge \citep{feltz} and the disk \citep{hay01}.

{\em iv)} The $v$ filter is strongly affected by two $CN$ molecular absorption 
bands ($\lambda=4142$, $\lambda=4215$ \AA). Stars with an over-abundance 
of carbon ($C$) and/or nitrogen ($N$), i.e. $CH$- and/or $CN$-strong stars, 
will have, at fixed color, a larger $m_1$ value, a fundamental property for 
identifying stars with different $CNO$ abundances in Globular Clusters 
(GCs, CA07, \citealt{io09, io11}).

On the other hand, the \strom system presents two relevant drawbacks. 

{\em i\/}) the $u$ and $v$ bands have short effective  
wavelengths, namely $\lambda_{eff}\,=\,3450$ and $\lambda_{eff}\,=\,4110\,$\AA. 
As a consequence the ability to perform accurate photometry 
with current CCD detectors is hampered by their reduced sensitivity 
in this wavelength region. 

{\em ii\/}) The intrinsic accuracy of the stellar parameters,  
estimated using \strom indices, strongly depends on 
the accuracy of the absolute zero-point calibrations. This 
typically means a precision better than 0.03 mag. This precision  
could be easily accomplished in the era of photoelectric photometry,
but it is not trivial effort in the modern age of CCDs.

The calibration of \strom photometric indices to obtain stellar metal 
abundances is not a new technique. Empirical calibrations based on such 
a method have been given by \citet{strom64, bond70, craw75, nissen81}. 
In these works, most of the stars adopted to perform the 
calibration are nearby ($d \lesssim$ 100 pc) F and early-type G dwarfs, 
with $\feh >$ -0.8. The adopted samples include a significant fraction of 
young and intermediate-age disk stars, and a minority of low-mass, old stars.
Moreover, these calibrations are based on differential indices $\delta_{m1}$ and 
$\delta_{c1}$, i.e. $\delta_{m1} = m_{1,Hyades} (\beta) - m_{1,star} (\beta) $, 
where $\beta$ stands for $H_{\beta}$ and $m_{1,Hyades} (\beta)$ is the standard
relation between $m_1$ and $\beta$ for the Hyades given by \citet{craw75} and 
by \citet{ols84}. The $\delta_{m1}$ can also be defined with \bmy as indipendent 
parameter, but as pointed out by Crawford, $\delta_{m1} (b-y)$ is less sensitive 
to metallicity than $\delta_{m1} \beta$, because the \bmy color is also affected 
by blanketing.

   \begin{figure*}
   \includegraphics[width=16truecm, height=10truecm]{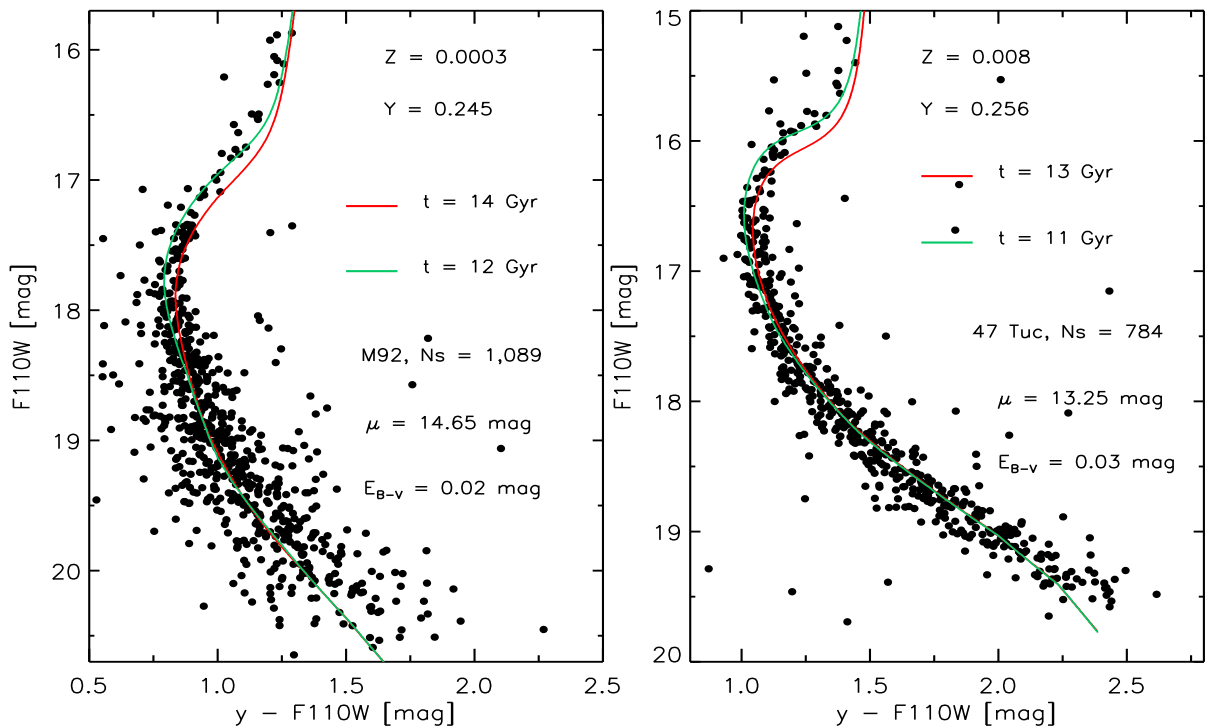}
      \caption{$F110W,\ (y-F110W)$ CMD for a metal-poor --M92 (left)-- and 
         a metal-rich --47~Tuc (right)-- globular. The green and red solid lines display cluster 
         isochrones. The adopted chemical 
         compositions, age,  true distance modulus and reddening are labeled.
         Tracks were computed by assuming $\alpha-$enhanced chemical mixtures (PI06)  
	 and transformed into the observational plane by adopting atmosphere models 
         with the same $\alpha-$enhancement.}
         \label{fig1}
   \end{figure*}

The differential indices $\delta_{m1}$ and $\delta_{c1}$ have the advantage 
that they measure mostly metallicity and surface gravity, respectively, and 
are free of temperature effects. However, they are affected by the 
uncertainty on the photometric zero-point of the Hyades standard 
relations and require accurate $\beta$ photometry.

\citet{ols84} derived a metal abundance calibration using high dispersion
spectroscopic measurements of \feh from \citet{cayrel83} and new \strom photometry 
for a sample of F, G and K dwarfs \citep{ols83}. 
Olsen provided a linear solution for $\delta_{m1} (b-y)$ in the range
-0.8 $ < \feh <$ 0.4 for F-G0 dwarfs, and a parabolic solution in the range 
-2.6 $ < \feh <$ 0.4 for G0-K1 dwarfs. However, only one calibrating star 
has $\feh <$ -1.9 and only three have $\feh <$ -1.5.

\citet[hereafter SN89]{schu89} performed new intrinsic color and metallicity 
calibrations based on a sample of 711 high-velocity and metal-poor stars 
with \strom photometry from \citet{schu88}. The stars have been selected to have 
spectral types in the range F0-K5, surface gravity 3.4 $ < log(g) < $ 5.4 and 
\feh abundances are based on high-resolution spectra of 
\citet{cayrel83, cayrel85, francois86}. In addition, these stars 
have $uvby$ photometry in the system of \citet{ols83, ols84}. The SN89 metallicity 
calibrations are based on the indices $m_1$ and $c_1$ and they are valid in the 
color range 0.22 $< (b-y)_0 <$ 0.59 mag and are reddening dependent.

\citet{hay02} claimed that systematic discrepancies of $\sim$ -0.1/0.3 dex
affected the SN89 photometric metallicity determinations of metal-rich stars 
in the quoted color range. They showed that this was a consequence of a 
mismatch between the standard sequence $m_1,\ (b - y)$ of the Hyades
used by SN89 to calibrate their metallicity scale, and the Olsen system. 
A new calibration was proposed by \citet{hay02}, on the basis of an enlarged 
spectroscopic data set, that makes the SN89 calibration applicable to the 
Olsen's photometric catalogs.

\citet{ramirez} proposed a new metallicity calibration for dwarf stars
based on an updated spectroscopic catalog by \citet{cayrel01} and 
the $m_1$ and $c_1$ indices, with different equations for three $b-y$ color ranges 
(0.19 $< \bmy <$ 0.35, 0.35 $< \bmy <$ 0.50, 0.50 $< \bmy <$ 0.80 mag).
These relations are valid over a broad metallicity range (-2.5 $ < \feh < $ 0.4) 
and for $log(g) > $ 3.4.

\citet[hereafter AR10]{arnadottir}, based on their new compilation of dwarf stars, 
tested some of the most recent and/or more popular metallicity calibrations 
\citep[SN89]{ols84,hay02,ramirez}. They found that the 
calibrations by SN89 and by \citet{ramirez} perform equally well, but the 
latter covers a larger parameter space. 
They suggested to adopt the calibration by \citet{ramirez} and the calibration
by \citet{ols84} for dwarfs redder than $\bmy =$ 0.8 mag.

More recently, \citet[hereafter CAS11]{casagrande} presented a new empirical 
metallicity calibration for dwarf stars based on the $m_1$ and $c_1$ indices,
covering the metallicity range -2.0 $ \lesssim \feh \lesssim$ 0.5 and 
0.23 $< \bmy <$ 0.63 mag. They validated the new calibrations by estimating the
metallicity of field dwarfs and open clusters, and showed that they give 
reliable abundance estimates with a dispersion smaller than 0.1 dex.

All these relations to estimate the metal abundance of dwarf stars are hampered 
by the presence of molecular $CN$,$CH$ and $NH$-bands that affect the \strom $uvb$ 
filters, and in turn the global metallicity estimates. Moreover, the quoted 
relations include the $c_1$ index, that is based on the $u$ filter. 
As already mentioned, observations in the $u$ filter are very demanding concerning 
the telescope time and the photometry in this band is less accurate due to the 
reduced CCD sensitivity in this wavelength region. 

We derive, for the first time, a theoretical calibration of a metallicity diagnostic 
based on the $m_1$ index and on visual--near-infrared (NIR) colors for dwarf 
stars. The visual--NIR ($y$, $J,H,K$) colors adopted in our new Metallicity-Index-Color 
(MIC) relations have two clear advantages when compared with optical colors: 
{\em i) \/} they are not hampered by the presence of $CN$,$CH$ and $NH$-bands;
{\em ii) \/} strong sensitivity to effective temperature. 
This means that the quoted molecular bands still affect the MIC relations, but 
only through the \strom $m_1$ index.

The new MIC relations are based on $\alpha$-enhanced evolutionary models and 
$\alpha$--enhanced bolometric corrections  and color-temperature transformations 
and are valid in the metallicity range -2.5 $ < \feh <$ 0.5, and for dwarf 
stars in the mass range 0.5 $< M < $ 0.85 $M_\odot$  (4.5 $< log(g) < $ 5).

The structure of the current paper is as follows. 
In \S 2 we discuss in detail the photometric catalogs of Galactic Globular clusters (GGCs) 
adopted to validate evolutionary models in the visual--NIR colors.
Section 3 deals with the approach adopted to calibrate the visual--NIR MIC relations, 
while in \S 4 we present the different tests we performed to validate the current 
theoretical calibrations together with the comparison between photometric estimates 
and spectroscopic measurements of metal abundances. We summarize the results and 
briefly discuss further improvements and applications of the new MIC relations 
in \S 5.


\section{Observations and data reduction}

We selected two GGCs, namely  M~92 (NGC~6341) and 47~Tuc (NGC~104),
to check the plausibility of the theoretical models we adopt to perform 
the metallicity calibration, since they cover a broad range of metal abundance 
(-2.31 $< \feh <$ -0.72, \citealt{harris}).

The \strom photometric catalog of M~92 (NGC~6341) adopted in this investigation 
was obtained with images collected with the 2.56m Nordic Optical Telescope (NOT) 
on La Palma \citep{gru00}, while the catalog for 47~Tuc (NGC~104) with images 
collected with the 1.54m Danish Telescope 
on La Silla (ESO, \citealt{gru02}).

Data for M~92 were collected during June 1998 and stars from the 
lists of \citet{ols83, ols84} and \citet{schu88} were observed 
to calibrate the instrumental $uvby$ magnitudes.
The total field of view (FoV) i $\sim 4 \times 4$ arcmin, 
including the cluster center.
The pixel scale is 0\farcs11 per pixel and the seeing ranges between 0\farcs5 and 1\farcs0~ 
(for more details see \citealt[CA07]{gru00}).

Data for 47~Tuc were collected during 10 nights in October 1997, using
the Danish Faint Object Spectrograph and Camera. The field 
of view covered by these data is approximately 11 arcmin across, excluding
the cluster center. The pixel scale is 0\farcs39 per pixel and the seeing 
ranges between 1\farcs3 and 2\farcs2. Approximately 150 different standard 
stars from the lists of \citet{ols83, ols84} and \citet{schu88} were also observed to
calibrate the data (for more details see \citealt[CA07]{gru02}).

We cross-correlated the \strom catalogs for the two Galactic globular 
clusters (GGCs) with the $F110W,F160W$ NIR photometry collected with 
the Wide Field Camera 3 (WFC3) on board of the Hubble 
Space Telescope. The log of the WFC3 images adopted in this investigation
is given in Table~1. The images were reduced by using a software package that is based largely
on the algorithms described by  \citep{anderson06}. 
Details on this program will be given in a stand-alone paper (Anderson et al.\ in preparation).
The data have been calibrated to Vega adopting the prescriptions by 
\citep{bedin05} and using the current on-line estimates for zero points and 
encircled energies\footnote{\textsf{http:\/\/www.stsci.edu\/hst\/wfc3\/phot\_zp\_lbn}}.
The final catalog for M~92 includes 7,219 stars with both NIR and \strom photometry,
covering a FoV of $\approx 2 \times 1$ arcmin. The catalog 
for 47~Tuc includes 12,227 stars overlapping with \strom photometry 
over a region of $\approx  3 \times 10$ arcmin at a distance of about 
12 arcminutes from the cluster center.

Fig.~1 shows the $F110W,\ y - F110W$ CMD for M~92 (left panel)
and 47~Tuc (right). Both catalogs are selected in photometric accuracy
and in distance from the cluster center.
The CMDs display the magnitudes fainter than $F110W \lesssim$ 15.5 (M~92) and 
$F110W \lesssim$ 15.0 mag (47~Tuc), since the WFC3 photometry for these clusters
is saturated at brighter magnitudes, and we are focusing our attention on main 
sequence stars.
In order to constrain the plausibility of the theoretical framework 
we adopt for the calibration of the new MIC relations for dwarf stars, 
we compare current visual-NIR photometry with evolutionary prescriptions. 
We adopted a true distance modulus of $\mu = 14.65$ mag
and a mean reddening of $E(B-V) = 0.02$ mag for M~92 \citep{dicecco}, 
and $\mu = 13.25$ mag and $E(B-V) = 0.03$ mag for 47~Tuc \citep{bono08}.  
The extinction coefficients are estimated by applying 
the \citet{card89} reddening relations and $R_V = A_V/E(B-V) = 3.1$, finding 
$A_{F110W} = 0.32 \times A_V$ and $E(y - F110W)= 2.12 \times E(B-V)$ mag.
The green and red solid lines in Fig.~1 show two isochrones with appropriate
ages and chemical compositions, namely  $Z$=0.0003, $Y$=0.245, $t$=12 and $t$ = 14 Gyr, 
and $Z$=0.008, $Y$=0.256, $t$=11 and $t$ = 13 Gyr for M~92 \citep{dicecco,brasseur} and 
47~Tuc \citep{bono08}, respectively. 
Isochrones are from the BASTI data base and are based on $\alpha$-enhanced 
($\afe=0.4$) evolutionary models \citep[hereafter PI06]{pietri06}.
Evolutionary prescriptions were transformed into the observational plane 
by using atmosphere models computed assuming $\alpha$-enhanced mixtures. 
Data plotted in Fig.~1 show that theory and observations, within the errors, 
agree quite well over the entire magnitude range for both clusters. 

\section{Calibration of new optical--NIR metallicity indices for dwarf stars}

Independent MIC relations were derived using cluster isochrones based on 
$\alpha$-enhanced evolutionary models (PI06).
Theoretical predictions were transformed into the observational plane
by adopting bolometric corrections (BCs) and Color--Temperature 
Relations (CTRs) based on atmosphere models computed assuming the same heavy 
element abundances \citep[PI06,][]{CK06}. 
The Vega flux adopted is from \citet{CK94}\footnote{The complete set of BCs, CTRs 
and the Vega flux are available at http://wwwuser.oat.ts.astro.it/castelli}.
The metallicities used for the calibration of the MIC relations are: 
Z = 0.0001, 0.0003, 0.0006, 0.001, 0.002, 0.004, 0.01, 0.02 and 0.03. 
The adopted Z values indicate the global abundance of heavy elements in the             
chemical mixture, with a solar metal abundance of ${(\zx)_\odot}=0.0245$. 

Fig.~2 shows the nine isochrones plotted in different visual--NIR (left) and \strom 
(right) MIC planes. 
{The $J,H,K$ filters where transformed into the 2MASS photometric system
by applying the color transformations by \citet{carpenter}.
Panels a), b) and c) display the $m_1$ index versus 
three visual-NIR colors ($y - K$, $y - H$, $y - J$), while the panels d), e) 
and f) the $m_1$ index versus three \strom colors ($u - y$, $v - y$, $b - y$). 
The evolutionary phases plotted in this figure range from approximately 
4 mag fainter ($M_d$, empty squares) than the  Main Sequence Turn-Off (MSTO) 
to 1 mag fainter ($M_u$, asterisks) than the MSTO. For a metal-intermediate 
chemical composition (Z=0.002, Y=0.248) the quoted limits imply absolute 
visual magnitudes of $M_V\approx$7.8 (M/M$_\odot$=0.5) and 
$M_V\approx$5.2 mag (M/M$_\odot$=0.75).

Data plotted in Fig.~2 show that the $m_1$ versus visual--NIR colors 
planes have a stronger sensitivity to metallicity of dwarf stars, 
when compared with the $m_1$ versus \strom colors. 
The two different sets of MIC relations cover similar $m_1$
values but the visual--NIR colors --panels a),b),c)-- show a stronger 
sensitivity in the faint magnitude limit and an almost linear change 
when moving from metal-poor to metal-rich stellar structures. 
The \strom colors --panels d),e),f)-- show a minimal sensitivity 
for stellar structures more metal-rich than $\mh\gtrsim$-1.0.  
Moreover and even more importantly, the slopes of the MIC relations 
based on visual--NIR colors are on average shallower than the 
MIC relations based on \strom colors. This means that the former 
indices have, at fixed $m_1$ value, a stronger temperature sensitivity.    

MIC relations for dwarf stars based on \strom colors are affected by 
the presence of molecular bands, such as $CN$,$CH$ and $NH$.
As a matter of fact, two strong cyanogen ($CN$) molecular absorption bands 
are located at $\lambda=4142$ and $\lambda=4215$ \AA, i.e. very close to 
the effective wavelength of the $v$ filter ($\lambda_{eff}=4110$,$\;\; 
\Delta \lambda=190$ \AA). Moreover, the strong $CH$ molecular band 
located in the Fraunhofer's $G-$band ($\lambda=4300$ \AA) might affect 
both the $v$ and the $b$ magnitude. It is noteworthy that the molecular 
$NH$ band  at $\lambda=3360$ \AA, and the two $CN$ bands at $\lambda=3590$ 
and $\lambda=3883$ \AA~ might affect the $u$ ($\lambda_{eff}=3450$,$\;\;
\Delta \lambda=300$ \AA) magnitude (see, e.g. Smith 1987).    
To decrease the contamination by molecular bands in the color index, 
we decided to adopt only colors based on the $y$-band and on the 
NIR bands in our new calibration of MIC relations. The main advantage 
of this approach is that the aforementioned molecular bands only 
affect the \strom $m_1$ index.

   \begin{figure*}
   \includegraphics[width=17truecm, height=17truecm]{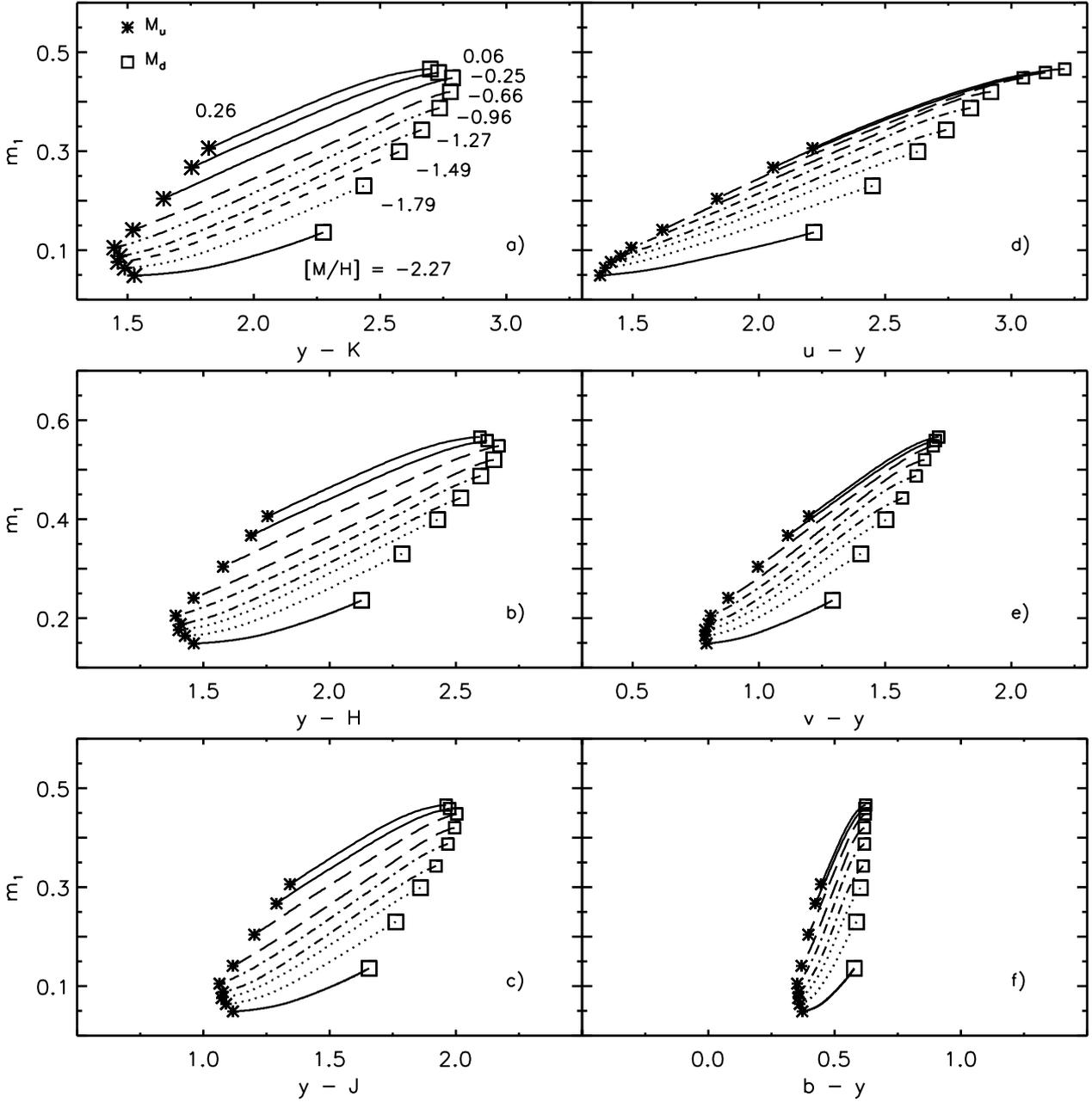} 
      \caption{Left panels -- $m_1$ vs $y-K$ plane [panel a)] for 
         isochrones at fixed cluster age ($t$=12 Gyr) and different global 
         metallicities ($\mh$, see labeled values). The evolutionary phases 
         range from $\approx$ 4 mag fainter 
         ($M_d$, empty squares) than the MSTO to $\approx$ 1 mag fainter 
         ($M_u$, asterisks) than the MSTO. Evolutionary tracks were computed 
         by assuming $\alpha-$enhanced chemical mixtures (PI06) and transformed 
	 into the observational plane by adopting atmosphere models computed 
         assuming the same $\alpha-$enhancements. The panels b) and c) similar 
         relations, but in the $m_1$ vs $y-H$ and in the $m_1$ vs $y-J$ plane. 
         Right panels -- Same as the left, but for the $m_1$ vs $u-y$ 
         [panel d)], $m_1$ vs $v-y$ [panel e)], $m_1$ vs $b-y$ [panel f)] 
         planes.
	     }
	 \label{fig2}
   \end{figure*}

We derived theoretical MIC relations based on $m_1$ and the $y$--NIR colors 
based on WFC3 ($F110W, F160W$) and 2MASS ($J,H,K$) bands. Together with the 
classical $m_1$ index, we also computed independent MIC relations for the 
reddening-free parameter $[m] = m_1\, +\, 0.3\, \times (\bmy)$, to overcome 
deceptive uncertainties caused by differential reddening.
To select the $m_1$ and the $[m]$ values along the individual isochrones 
we followed the same approach adopted in CA07.  A multilinear regression fit 
was performed to estimate the coefficients of the MIC relations for the 
$m_1$ and the $[m]$ indices as a function of the five $CI$s, namely 
$y - F110W$, $y - F160W$, $y - J$, $y - H$ and $y - K$:

\begin{eqnarray*}
m_1 = \alpha\, + \beta\,\mh + \gamma\, CI +
\delta\, (CI \times m_1) + \epsilon\, CI^2 + \\
\zeta\, m_1^2 + \eta\, (CI^2 \times m_1) + \theta\, (CI \times m_1^2) + 
\iota\, (CI^2 \times m_1^2) + \\
\kappa\, (CI \times \mh) + 
\lambda\, (m_1 \times \mh)
\end{eqnarray*}

where the symbols have their usual meaning. 
The adoption of eleven terms, compared to the four terms of the 
$m_1$ versus \strom color calibration for red giants, is due to the nonlinearity 
of the $m_1$ versus visual--NIR color relations for dwarf stars. 
To select the form of the analytical relation we followed the forms
adopted for the $m_1$ and the $hk$ metallicity index calibrations in 
CA07 and \citet{io11}, respectively. 
We performed several tests finding the best solution of the 
multilinear regression fit when adopting the $m_1$ index as an independent variable.  
We then selected the solution with the lowest chi-square of
the multilinear regression fit. As a further check, we estimated the Root Mean Square (RMS)
deviations of the fitted points from the fit and the values range between
0.0004 to 0.0012. Moreover, the multi-correlation parameters attain values close to 1.
The coefficients of the fits, together with their uncertainties, the 
for the ten MIC relations, are listed in Table~2. The RMS values and the 
multi-correlation parameters of the different relations are listed in the 
last tow columns.

The above MIC relations are valid in the following color ranges,
0.0 $< m_1 <$ 0.6 mag, 0.1 $< [m] <$ 0.8 mag,
1.1 $< y - F110W <$ 1.9 mag, 1.45 $< y - F160W <$ 2.6 mag,
1.1 $< y - J <$ 2.0 mag, 1.1 $< y - H <$ 2.5 mag, and
1.5 $< y - K <$ 2.8 mag, and for dwarf stars in the mass range 
0.5 $\lesssim M \lesssim $ 0.85 $M_\odot$ (4.5 $< log(g) < $ 5).

\section{Validation of the new metallicity calibration}\label{validation}
\subsection{Field dwarf stars}
In order to validate the new theoretical calibration of the $m_1$ index 
based on $y$--NIR colors for dwarf stars we estimate the metallicity of 
field dwarfs for which {\it vby\/} photometry, NIR photometry and 
high-resolution spectroscopy are available.
We selected two catalogs from the literature, the former by AR10 includes 
451 dwarfs, while the latter by CAS11 includes 1,498 dwarfs.

The {\it vby\/} photometry for the stars in the AR10 catalog was retrieved
by different studies and transformed into the \strom system of \citet{ols93}, 
while the high-resolution spectroscopic abundances were collected
from different analysis and homogenized to the spectroscopic system of 
\citet{valentifish}.
The extinction values were estimated by using the reddening map
by \citet{schlegel98} and modeling the dust in the Galactic disk 
with a thin exponential disk with a scale-height of
$\approx$ 125 pc (see AR10 for more details). Stars with a reddening
$E(B-V) <$ 0.02 mag are assumed to be unreddened and no correction
was applied to the photometry of these stars (AR10).

To unredden the $m_1$ index and the colors of the dwarf stars 
we adopt $E(m_1) = - 0.30 \times E(\bmy)$ \citep{io07}, and 
$E(y - J)= 2.23 \times E(B-V)$, 
$E(y - H)= 2.56 \times E(B-V)$
$E(y - K)= 2.75 \times E(B-V)$, estimated assuming the \citet{card89} 
reddening relation and $R_V = A_V/E(B-V) = 3.1$.

This sample was cross-correlated with the 2MASS photometric 
catalog  by retrieving for each dwarf star $J,H,K$-band measurements.

   \begin{figure}
   \includegraphics[width=0.5\textwidth]{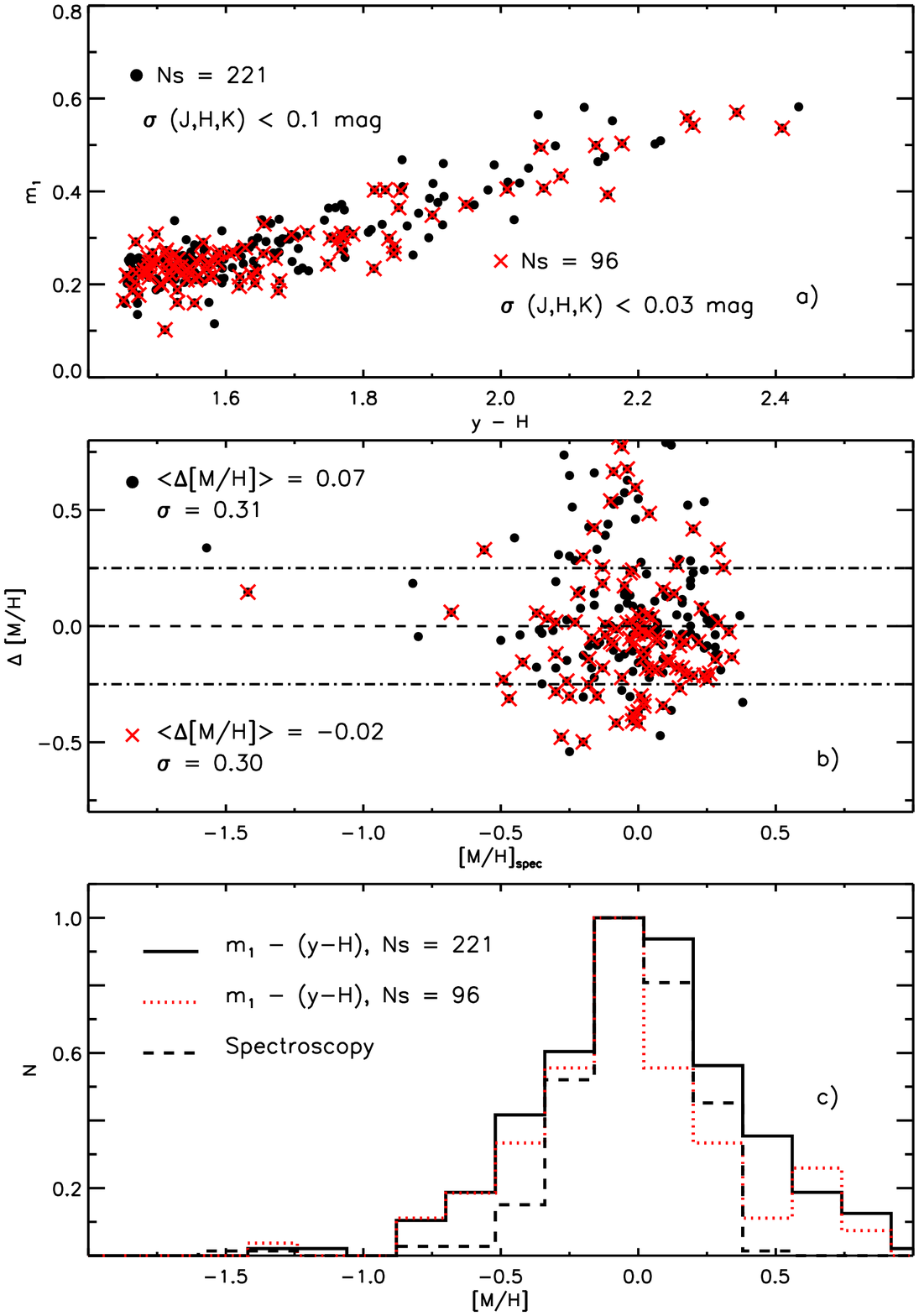}
      \caption{Panel a): selected field dwarf stars
       from the sample of AR10 plotted in the unreddened $m_1 ,\ y - H$ 
       plane (Ns = 221, filled dots). 
       Stars selected for $\sigma_{J,H,K} <$ 0.03 mag are marked with red crosses (Ns = 96).
       Panel b): Difference between photometric and spectroscopic metallicities,
       $\Delta \mh = (\mh_{\hbox{\footnotesize phot}} - \mh_{\hbox{\footnotesize spec}})$, 
       plotted versus
       $\mh_{\hbox{\footnotesize spec}}$ for the 221 field dwarfs (filled dots).
       Photometric metallicities are based on the $m_1, \, y - H$ MIC relation.
       Panel c): photometric metallicity distribution for the 221 dwarfs obtained with
       the $m_1, \, y - H$ MIC relation (black solid line), compared to the same 
       distribution but for the selected 96 stars (red dotted) and to the spectroscopic
       distribution (black dashed).
            }
	 \label{fig3}
   \end{figure}

We selected, from the AR10 sample, dwarf stars with a measurement of the 
global metallicity $\mh$ from \citet{valentifish} and with unreddened colors 
falling inside the color range of current calibrations. The metallicity range 
covered by our MIC relations is $-2.3 < \mh < 0.3$, but we select stars with 
$-2.5 < \mh < 0.5$ to account for uncertainties in spectroscopic abundances 
and in the metallicity scale \citep{kra03}. We further select stars in 
photometric accuracy ($\sigma_{J,H,K} \le 0.1$ mag), ending up with a sample of 
221 field dwarfs. The selected stars are plotted in the $m_1 ,\ y - H$ plane 
in panel a) of Fig.~3 (filled dots).  
The observed spread is mainly due to photometric errors and to the 
uncertainty in the calibration equation, which account for $\approx$ 0.25 dex, 
and to spectroscopic measurement errors. Panel b) of the same figure shows the 
difference between the photometric and the spectroscopic metallicity 
($\Delta \mh = (\mh_{\hbox{\footnotesize phot}} - \mh_{\hbox{\footnotesize spec}})$ 
for the 221 field dwarf stars as a function of their spectroscopic metal abundances 
($\mh_{spec}$). Data plotted in this panel show that there is a group of stars 
with photometric abundances systematically more metal-rich than the 
spectroscopic ones. 
To constrain the nature of this drift, we adopted more restrictive selection 
criteria for the photometry. By selecting stars with $\sigma_{J,H,K} \le 0.03$ mag, 
we ended up with a sample of 96 dwarfs (red asterisks), but the outliers almost 
completely disappear. 
The mean difference between photometric and spectroscopic abundances was 
estimated by adopting the biweight algorithm and for $m_1, \ y-H$ MIC relation 
it is -0.02$\pm$0.02 dex, with $\sigma$ = 0.30 dex.
The same difference, but based on the entire sample of 221 dwarfs is 
0.07$\pm$0.02 dex, with $\sigma$ = 0.31 dex. 
Panel c) of Fig.~3 shows the comparison between the photometric 
metallicity distribution based on the $m_1,\ y-H$ MIC relation (solid line) 
for the 221 dwarfs compared to the distribution obtained for the selected 96 dwarfs 
(red dotted) and the spectroscopic metallicity distribution (dashed).
Data indicate that spectroscopic and photometric metallicity distributions 
agree quite well within an intrinsic dispersion of the order of 0.3 dex. 
Photometric metallicity distributions estimated by adopting the other MIC relations
agree with each other and the mean difference between photometric and 
spectroscopic abundances for the 96 stars derived averaging the $m_1,\ y-J$, 
$m_1,\ y-H$, $m_1,\ y-K$ relations is $-0.02\pm0.10$ dex, with a mean intrinsic
dispersion of $\sigma =$0.31 dex. The difference attains similar values 
--$0.07+0.06$ dex, with $\sigma =$0.31 dex-- by averaging the MIC relations 
based on the reddening free metallicity indices ($[m],\ y-J$, $[m],\ y-H$, $[m],\ y-K$).

   \begin{figure}
   \includegraphics[width=0.5\textwidth]{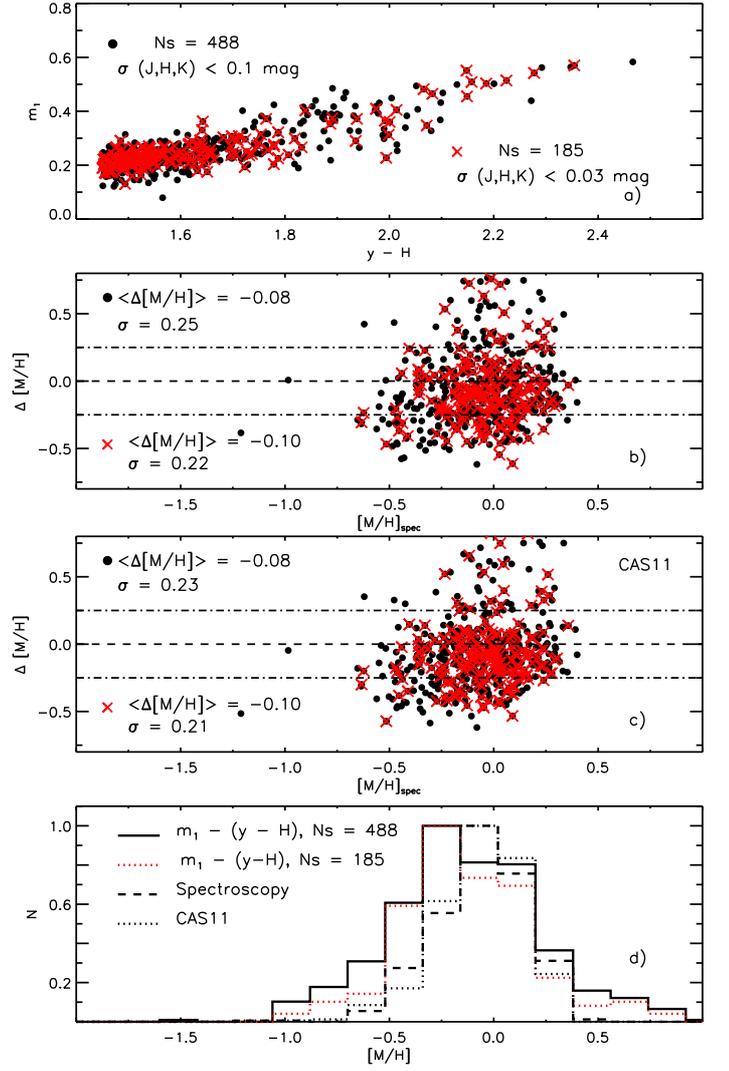}
      \caption{Panel a): selected field dwarf stars
       from the sample of CAS11 plotted in the unreddened $m_1 ,\ y - H$ plane (Ns =  488, filled dots). 
       Stars selected for $\sigma_{J,H,K} <$ 0.03 mag are marked with red crosses (Ns = 185).
       Panel b): Difference between photometric and spectroscopic metallicities,
       plotted versus the spectroscopic metallicity for the 488 field dwarfs.
       Photometric metallicities are based on the $m_1, \, y - H$ MIC relation.
       Panel c): Difference between photometric metallicities based on 
       our $m_1, \, y - H$ MIC relation and photometric metallicities by CAS11
       plotted versus spectroscopic metallicity.
       Panel d): photometric metallicity distributions for the 488 dwarfs (black solid line) 
       and for the selected 185 stars (red dotted) obtained with
       the $m_1, \, y - H$ MIC relation,§ compared to the photometric 
       distribution by CAS11 (black dotted) and to the spectroscopic
       distribution (black dashed).
	     }
	 \label{fig4}
   \end{figure}

To further validate the new optical--NIR metallicity calibrations, we also adopt 
the sample by CAS11.
The \strom {\it vby\/} photometry for the dwarf stars was retrieved
by different studies and transformed into the photometric system of 
\citet{ols93} by Nordstrom et al. (2004). The interested reader is 
referred to this paper for more details concerning the sample selection.
The high-resolution spectroscopic abundances were collected by CAS11 
from three large surveys, namely \citet{valentifish, sousa08, bensby11}. 
These measurements are consistent with each other and with the 
homogenized sample of AR10. The final sample includes 1,498 dwarf stars 
with \strom photometry and high-resolution spectroscopic measurements.
This sample was cross-correlated with the 2MASS catalog and we ended up 
with a sample of 1,135 stars with $J,H,K$-band photometry.
Reddening corrections were applied only to stars with distances larger 
than 40 pc and with $E(b-y) >$ 0.01 mag, otherwise the extinction was 
assumed to be vanishing (CAS11).
For our analysis we only selected  dwarf stars with $log (g) >$ 3.5, 
$-2.5 < \mh < 0.5$ and in the color range of validity of our metallicity 
calibration. We further select the sample in photometric accuracy, i.e. 
$\sigma_{J,H,K} \le$ 0.1 mag, ending up with 488 dwarfs.
Panel a) of Fig.~4 shows the selected stars in the $m_1,\ y - H$ plane,
while panel b) shows the difference between photometric 
and spectroscopic metallicity as a function of the spectroscopic 
metal abundances ($\mh_{spec}$) for the 488 dwarfs (filled dots). 
Photometric metallicities are estimated by adopting the $m_1 ,\ y-H$ MIC relation.
The observed spread is mainly due to photometric errors and to the uncertainty 
in the calibration equation, which account for $\approx$ 0.2 dex, and to spectroscopic 
measurement errors. The spectroscopic uncertainty has a mean value of about $0.05$ dex, and 
the difference between metallicities derived in the three spectroscopic surveys 
adopted ranges from $0.03$ to $0.05$ dex with a dispersion of $\sim$ 0.05 dex
(see CAS11 and \citealt{valentifish} for more details).

As in the case of AR10 sample, there is a group of stars with photometric 
metallicities systematically more metal-rich than spectroscopic abundances.
Most of them disappear once we apply more restrictive selection criteria 
in photometric accuracy, $\sigma_{J,H,K} \le$ 0.03 mag, ending up with a 
sample of 185 dwarfs (red asterisks).
On the other hand, the figure shows that for both samples a small shift of 
photometric metallicities being more metal-poor than spectroscopic measurements 
is present. The mean difference 
between photometric and spectroscopic metallicities
estimated by using the biweight algorithm for the 185 stars
is -0.10$\pm$0.02 dex, with $\sigma$ = 0.22 dex, while for 
the 488 stars is -0.08$\pm$0.02 dex, with $\sigma$ = 0.25 dex. 
It is noteworthy that in spite of the shift the shape of the photometric metallicity 
distributions --solid and red dotted lines in panel d)-- agrees quite well with 
the shape of the spectroscopic one (dashed). 
A culprit for the systematic shift between photometric and spectroscopic
abundances might be an $\alpha$-enhancement for field dwarfs smaller 
than the assumed $\alpha$ value.
The evolutionary models adopted to perform our theoretical metallicity calibration 
have been computed assuming $[\alpha/Fe]$ = 0.4. This is the typical enhancement 
found in cluster stars using high-resolution spectra (Kraft et al. Gratton et al. 
old and new). On the other hand, dwarfs in CAS11 sample have $[\alpha/Fe] <$ 0.4, 
with a median value of $[\alpha/Fe] \sim$ 0.05 (see their Fig.~9).
The $\alpha$-enhanced isochrones have typically redder colors in the 
$m_1$ versus $CI$ planes compared with scaled-solar models (see Fig.~16 in CA07). 
Therefore, $\alpha$-enhanced theoretical MIC relations give more metal-poor 
metallicity estimates for dwarf stars that are less $\alpha$-enhanced than 
cluster stars.
This effect decreases when adopting the $m_1, y-K$ relation, 
obtaining a mean difference of 0.0 dex with a dispersion of $\sigma$ = 0.25 dex for the
185 dwarfs, and of 0.03 dex and $\sigma$ = 0.32 dex for the 488 dwarfs.
Panel c) of Fig.~4 shows the difference between photometric metallicities 
estimated by adopting our MIC relation $m_1,\ y - H$ and photometric 
metallicities by CAS11 plotted versus the spectroscopic metal abundances for 
the two sample of dwarfs. Note that CAS11 estimated photometric \feh by adopting 
their fully empirical calibration (see Equations (2) or (3) of their paper), 
and the global metallicity \mh by using a \strom index ($a_1$) to estimate 
a proxy of the $[\alpha/Fe]$ abundance of the stars. The mean difference 
between the photometric metallicities is -0.10$\pm$0.02 dex, with a 
dispersion of $\sigma$ = 0.21 dex for the 185 stars, and -0.08$\pm$0.02 dex, 
with $\sigma$ = 0.23 dex for the 488 stars.
CAS11 metallicity distribution for the sample of 488 dwarfs is also showed 
in panel d) of Fig.~4.
Photometric metallicity estimates obtained by using the other MIC relations 
agree very well with each other, and the biweight mean difference with the 
spectroscopic measurements for the 185 stars derived averaging the $m_1,\ y-J$, 
$m_1,\ y-H$, $m_1,\ y-K$ relations is $-0.06\pm0.07$ dex, with 
a mean intrinsic dispersion of $\sigma$ = 0.22 dex, while is 
$-0.01\pm0.06$ dex, with $\sigma =$0.25 dex, by averaging the $[m],\ y-J$, 
$[m],\ y-H$, $[m],\ y-K$ relations.

In order to constrain the possible dependence on the adopted theoretical framework, 
we used the evolutionary models publicly available in the Dartmouth database 
\citep{dotter07,dotter08}. The cluster isochrones are transformed into the 
observational plane by using the semi-empirical CTRs
by \citet{clem} for the \strom colors and the CTRs 
predicted by PHOENIX atmosphere models for the 2MASS colors \citep{hausa,hausb}.
 
We followed the same approach adopted to calibrate the BASTI models. In particular, 
we selected the same metallicity range, i.e. 0.0001$\lesssim$ Z $ \lesssim$ 0.03.
Six Dartmouth sets of isochrones are available in this metallicity range, i.e. 
Z = 0.0001, 0.000354, 0.001, 0.00354, 0.01, 0.0352. Note that to account for the fact 
that the Dartmouth isochrones include gravitational settling of heavy elements, 
we selected cluster isochrones of 10.5 Gyr.   
We derived the $m_1 ,\ y - H$ MIC relation and we applied it to estimate 
the metallicities of the CAS11 field dwarf sample. 
The difference between photometric and spectroscopic metallicities, estimated by 
adopting the biweight algorithm, is 0.20 dex with $\sigma$= 0.21 dex.
Thus suggesting that the metallicity estimates based on the Dartmouth models are 
slightly more metal-rich than spectroscopic measurements.

The comparison between the different sets of isochrones have have already been 
discussed in the literature \citep{dotter07,dotter08,pietri06,pietri09}. 
We compared the two different sets of isochrones in the $m_1 ,\ y - H$ 
color--color plane and we found that the Dartmouth models appear, at 
fixed $y - H$ color, slightly bluer than the BASTI models. The difference might be due 
to the different sets of CTRs adopted and to the fact that the MIC relations based 
on the latter set relies on a finer metallicity grid (nine vs six isochrones).

\subsection{Cluster dwarf stars}

In order to validate the metallicity calibration based on the
visual--WFC3 NIR colors we adopt cluster dwarf stars.  
We selected two GGCs, namely M~92 and 47~Tuc, with 
$\feh$ = -2.31 and $\feh$ = -0.72, respectively \citep{harris}. 

The use of cluster data brings forward three indisputable advantages: 
{\em a)}-- the evolutionary status (age, effective temperature, surface 
gravity, stellar mass) of cluster dwarfs is well established;
{\em b)}-- the abundance of iron and $\alpha$-elements is known with 
high-precision;  
{\em b)}-- the selected clusters cover a broad range in iron abundance. 

Data for these clusters have already been presented in \S 2.  
The top panel of Fig.~5 shows 47~Tuc MS stars plotted in the $m_1,\ y - F160W$
plane. Stars are selected in magnitude, 18.5 $< y <$ 20.5 mag, and for the color 
ranges of validity of our metallicity calibration, i.e. 1.45 $< (y - F160W)_0 <$ 2.6 mag,
0.0 $< m_{10} <$ 0.6 mag. A further selection is performed in photometric
accuracy, $\sigma_y <$ 0.05 mag, and in distance from the cluster center,
$RA <$ 5.4 $\deg$. The final sample includes 111 stars, covering a 
region of 47~Tuc between 5.3 $< RA <$ 5.4 $\deg$. 

   \begin{figure}
   \includegraphics[width=0.5\textwidth]{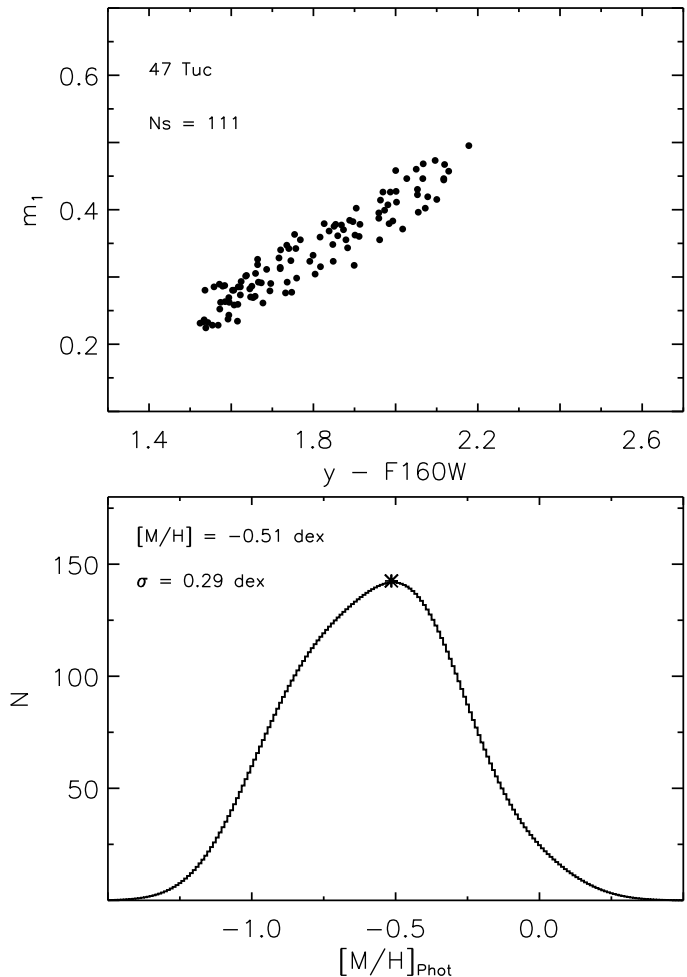} 
      \caption{Top: selected MS stars from the GGC 47~Tuc plotted in the 
         $m_1, \ y - F160W$ plane.
	 Bottom: photometric metallicity distribution obtained applying the 
	 $m_1, \ y - F160W$ MIC relation for the sample of 47~Tuc MS stars. }
	 \label{fig5}
   \end{figure}
   
The metallicity distribution of selected 47~Tuc MS stars obtained by applying 
the $m_1 , \ y - F160W$ MIC relation is showed in the bottom panel of Fig.~5.
The distribution has been smoothed by applying a Gaussian kernel having 
a standard deviation equal to the photometric error in the $m_1$ index, 
following the prescriptions of \citet{io09}.
The photometric metallicity distribution is clearly asymmetric and covers more 
than 1 dex in metal abundance. 
Fitting the distribution with a Gaussian we found a main peak around 
$\mh$ = -0.51 ($\feh$ = -0.86), with a dispersion of $\sigma$ = 0.29 dex. 
This estimate is in very good agreement with spectroscopic estimates of 
47~Tuc metal abundance from the literature ($\feh$  = -0.71$\pm$0.05, 
\citealt{rutledge} and $\feh$ = -0.76$\pm$ 0.02, \citealt{carretta09}).
The metallicity distributions obtained by applying the other MIC relations
agree with each other, with averaged mean peaks of -0.47$\pm$ 0.05 dex ($\feh=$ -0.82) and a
mean intrinsic dispersion of $\sigma =$ 0.30 dex ($m_1,\ y-F110W$, 
$m_1,\ y-F160W$ relations) and of -0.47$\pm$ 0.02 dex, 
with $\sigma =$ 0.28 dex ($[m],\ y-F110W$, $[m],\ y-F160W$ relations).

The large spread in the metallicity distributions might be due to photometric 
errors. The mean $y-F160W$ and $m_1$ color errors in the selected magnitude bin 
(18.5 $< y <$ 20.5 mag) are $\approx$ 0.015 and 0.025 mag, respectively.
We simulated the observed $m_1,\ y-F160W$ 47~Tuc color--color plane by
assuming isochrones with $Z$=0.008, $Y$=0.256, $t$=11 Gyr (see Section 1)
in the \strom and WFC3 NIR colors. We selected only points in the color ranges of
validity of the calibration, i.e. 1.45 $< (y - F160W) <$ 2.6 mag,
0.0 $< m_1 <$ 0.6 mag, and added to each point a random error 
drawn from a gaussian distribution with sigma equal to the photometric mean
error in the selected colors. The metallicities have been estimated from the
simulated dwarf sequence and we get a symmetric distribution with a peak
at $\mh =$ -0.45 and $\sigma$ = 0.30 dex.

To constrain the possible culprit for the asymmetry in the cluster metallicity distribution 
we select stars more metal-poor than $\mh <$ -0.8 (Ns = 21) according 
to the distribution obtained by applying the $m_1,\ y-F160W$ 
relation (Fig.~5), and we check for the presence of spatial distribution trends.
We compare the spatial distribution of selected stars with the rest of 
the sample ($\mh \ge$ -0.8, Ns = 90) and we do not detect any peculiarity. 

 \citet{anderson09} and \citet{milone12} have recently disclosed, based on HST data, 
 the presence of at least two sequences along 47~Tuc MS, corresponding to two different 
stellar populations with the same iron content, but probably different 
$He$ and light element abundances. The two stellar populations, defined as 
"first" and "second" generation have different spatial distributions. 
The $He/N$-enriched subsample, i.e. the second generation, is more centrally 
concentrated and accounts up to $\sim$70\% of the cluster population. 
The photometric metallicity distributions we obtain for the sample of
111 dwarfs of 47~Tuc seem to suggest the presence of stars more metal-poor 
than the mean cluster metallicity, but our sample is too small to draw firm conclusions. 

   \begin{figure}
   \includegraphics[width=0.5\textwidth]{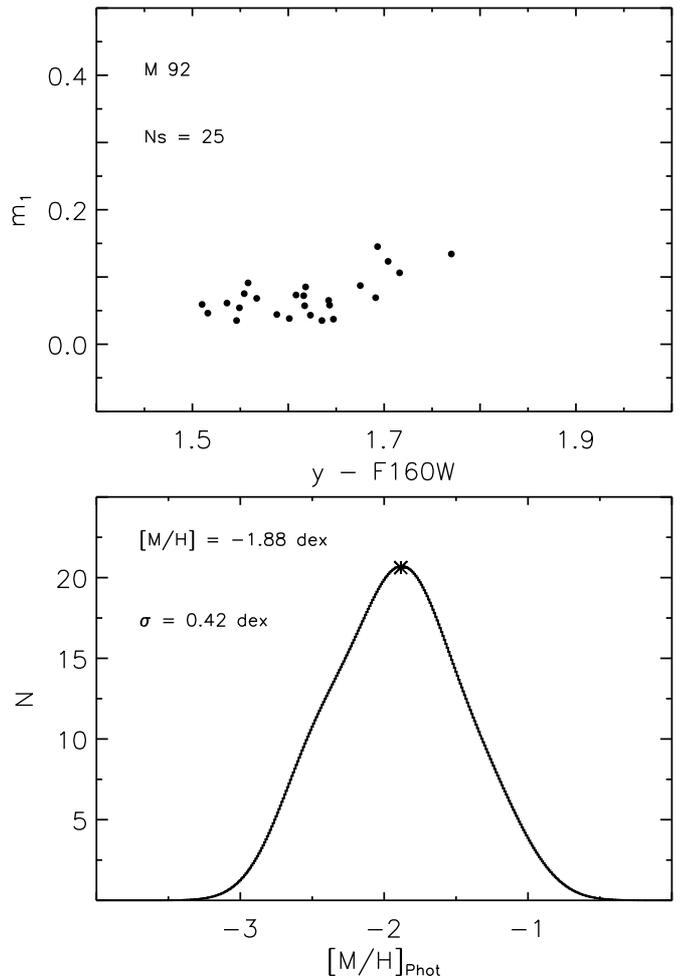}
      \caption{Same ad Fig.~5 but for the GGC M~92.}
	 \label{fig6}
   \end{figure}

The top panel of Fig.~6 shows M~92 selected MS stars plotted in the 
$m_1 , \ y - F160W$ plane. The sample is selected in magnitude, 
20.0 $< y <$ 20.7 mag, in photometric accuracy, $Sharpness <$ 0.8, 
and for the color range of validity of our metallicity calibration, ending
up with 25 stars.
The bottom panel of Fig.~6 shows the smoothed photometric metallicity distribution
obtained by applying the $m_1, \ y - F160W$ MIC relation to the sample.
The distribution covers more than 2 dex in metallicity and it has an asymmetric
shape although less pronounced that in the case of 47~Tuc.
By fitting the distribution with a Gaussian we obtain a main peak
at $\mh = -1.88$ ($\feh =$ -2.23) with a dispersion of $\sigma =$ 0.42 dex. 
The metallicity distributions obtained by applying the other MIC relations
agree with each other, with averaged mean peaks of -1.85$\pm$0.05 dex ($\feh =$ -2.20) 
and a mean intrinsic dispersion of $\sigma =$ 0.43 dex ($m_1,\ y-F110W$, 
$m_1,\ y-F160W$ relations) and of -2.17$\pm$0.20 dex ($\feh =$ -2.52), 
with $\sigma =$  0.35 dex ($[m],\ y-F110W$, $[m],\ y-F160W$ relations).
These estimate are in good agreement, within uncertainties, with spectroscopic 
estimates from the literature ($\feh =$ -2.24$\pm$0.10 \citealt{zinn} and 
$\feh =$ -2.35$\pm$0.05 \citealt{carretta09}).

M~92 shows the typical variations in $[C/Fe]$ and $[N/Fe]$ 
\citep{carbon82,langer,bellman}, together with the usual anticorrelations 
of most GGCs \citep{pila, sneden, kra94}.
Moreover, preliminary analysis of HST data suggest that also M~92 exhibits 
multiple sequences in the CMD in close analogy with 47~Tuc (Milone et al., in preparation).
On the other hand, most of the spread of the metallicity distribution is 
due to the large photometric errors of the catalog in this region of 
the CMD (see also Fig.~1). The mean $y-F160W$  and $m_1$ color errors 
in the selected magnitude bin (20.0 $< y <$ 20.7 mag) are 0.035 mag and
0.07, with maximum values of 0.11 mag and 0.13 mag, respectively.
As in the case of 47~Tuc, we simulated the observed $m_1,\ y-F160W$ color--color plane by
assuming isochrones with $Z$=0.0003, $Y$=0.245, $t$=13 Gyr (see Section 1)
in the \strom and WFC3 NIR colors. We selected only points in the color ranges of
validity of the calibration, i.e. 1.45 $< (y - F160W)_0 <$ 2.6 mag,
0.0 $< m_{10} <$ 0.6 mag, and added a random error to each magnitude
drawn from a gaussian distribution with sigma equal to the photometric mean
error in that band. The metallicities have been estimated from the
simulated dwarf sequence and we get a distribution with a large spread, with a peak
at $\mh =$ -1.50 and $\sigma$ = 0.45 dex (the adopted isochrones are slightly more metal-rich
than M~92).

Therefore the photometric error and the small sample does not allow us to
infer the presence of stars with different light-element abundances in the cluster.

\begin{table*}
\caption{Log of the NIR images collected with WFC3 on the HST for the Galactic globular cluster
47~Tuc (Program GO-11677, PI: H. Richer, 11453, PI: B. Hilbert) and M~92 (Program GO-1664, PI: T.M. Brown).}           
\label{table:1}      
\begin{tabular}{l c c c c }        
\hline\hline                 
Date & Exposure time & Filter & RA & DEC \\    
     &  (s)          &        & (h) & (deg) \\  
\hline                        
47~Tuc\\
Mar  3 2010       & $62+174+2\time1399$                          & F110W & 00:21:48 & -72:08:36 \\ 
Mar  3 2010       & $249+3\time299+4\time1199$                 & F160W & 00:21:48 & -72:08:36 \\ 
Apr  4 2010       & $102+174+2\time1399$                         & F110W & 00:21:29 & -72:06:45 \\ 
Apr  4 2010       & $4\time299+4\time1199$                       & F160W & 00:21:29 & -72:06:45 \\ 
Jun 12 2010       & $174+2\time1399$                               & F110W & 00:21:23 & -72:02:36 \\ 
Jun 12 2010       & $2\time149+2\time299+4\time1199$         & F160W & 00:21:23 & -72:02:36 \\ 
Jun 18 2010       & $102+174+2\time1399$                         & F110W & 00:21:39 & -72:00:14 \\ 
Jun 18 2010       & $4\time299+4\time1199$                       & F160W & 00:21:39 & -72:00:14 \\ 
Sep 19 2010       & $102+174+2\time1399$                         & F110W & 00:23:11 & -71:58:40 \\ 
Sep 19 2010       & $4\time299+4\time1199$                       & F160W & 00:23:11 & -71:58:40 \\ 
Jul 16,17,23 2009 & $18\time149$                                     & F110W & 00:22:38 & -72:04:04 \\ 
Jul 16,17,23 2009 & $42\time274$                                     & F110W & 00:22:38 & -72:04:04 \\ 
Aug 05 2010       & $102+174+2\time1399$                         & F110W & 00:22:20 & -71:58:18 \\ 
Aug 05 2010       & $4\time299+4\time1199$                       & F160W & 00:22:20 & -71:58:18 \\ 
Aug 14,15 2010    & $102+174+2\time1399$                         & F110W & 00:22:46 & -71:58:09 \\ 
Aug 14,15 2010    & $4\time299+4\time1199$                       & F160W & 00:22:46 & -71:58:09 \\ 
Mar 4 2010        & $102+174+2\time1399$                         & F110W & 00:22:09 & -72:09:34 \\ 
Mar 4 2010        & $4\time299+4\time1199$                       & F160W & 00:22:09 & -72:09:34 \\ 
Jul 29 2010       & $99+174+1199+1399$                           & F110W & 00:21:57 & -71:59:01 \\ 
Jul 29 2010       & $99+124+299+349+4\time1199$              & F160W & 00:21:57 & -71:59:01 \\ 
Oct 1 2010        & $102+174+2\time1399$                         & F110W & 00:23:32 & -71:59:47 \\ 
Oct 1 2010        & $4\time299+4\time1199$                       & F160W & 00:23:32 & -71:59:47 \\ 
Jan 15-28 2010    & $3\time32+41+102+21\time174+19\time299+37\time1199+40\time1399$ & F110W & 00:23:12 & -72:09:25 \\
Jan 15-28 2010    & $7\time32+41+14\time174+4\time299+9\time1199+29\time1399$ & F160W & 00:23:12 & -72:09:25 \\
May 3 2010        & $102+174+2\time1399$                         & F110W & 00:21:21 & -72:04:40 \\ 
May 3 2010        & $4\time299+4\time1199$                       & F160W & 00:21:21 & -72:04:40 \\ 
\hline                                   
\hline                                   
M~92\\
Oct 10 2009        & $2\time4+3\time49+299+2\time399$    & F110W & 17:17:07 & 43:07:58 \\
Oct 10 2009        & $2\time4+3\time49+299+2\time399$    & F160W & 17:17:07 & 43:07:58 \\
\hline                                   
\end{tabular}
\end{table*}

\begin{table*}
\caption{Multilinear regression coefficients for the \strom  
metallicity index:$m_1 = \alpha\, + \beta\,\feh + \gamma\, CI +
\delta\, (CI \times m_1) + \epsilon\, CI^2 + 
\zeta\, m_1^2 + \eta\, (CI^2 \times m_1) + \theta\, (CI \times m_1^2) + 
\iota\, (CI^2 \times m_1^2) + 
\kappa\, (CI \times \feh) + 
\lambda\, (m_1 \times \feh)$}             
\label{table:2}      
\begin{tabular}{l c c c c c c c c c c c c c}        
\hline\hline                 
Relation & $\alpha$ & $\beta$ & $\gamma$ & $\delta$ & $\epsilon$ & $\zeta$ & $\eta$ & $\theta$ & $\iota$ & $\kappa$ & $\lambda$ & Multicorr & RMS \\    
\hline                        
$m_1,  y - F110W$ & 0.015 & 0.027 & 0.066 & -0.073 & 0.731 & -0.027 & 0.069 & 1.289 & 0.535 & -0.233 & -1.780 & 1.000 & 0.0005    \\    
Error             & 0.007 & 0.002 & 0.016 & 0.009  & 0.111 & 0.002  & 0.006 & 0.017 & 0.031 & 0.020  & 0.080  & $(\ldots)$ & $(\ldots)$  \\    
$[m],  y - F110W$ & 0.100 & 0.027 & -0.048 & -0.045 & 0.559 & -0.027 & 0.030 & 1.412 & 0.434 & -0.365 & -1.220 & 1.000  & 0.0004    \\   
Error             & 0.007 & 0.002 & 0.018 & 0.012  & 0.077 & 0.002  & 0.005 & 0.031 & 0.017 & 0.020  & 0.042  & $(\ldots)$ & $(\ldots)$  \\
$m_1,  y - F160W$ & 0.026 & 0.031 & 0.029 & -0.027 & 0.963 & -0.019 & 0.030 & 0.913 & 0.302 & -0.162 & -1.239 & 1.001  & 0.0004     \\   
Error             & 0.007 & 0.002 & 0.011 & 0.004  & 0.115 & 0.001  & 0.006 & 0.019 & 0.015 & 0.010  & 0.060  & $(\ldots)$ & $(\ldots)$  \\
$[m],  y - F160W$ & 0.075 & 0.026 & 0.007 & -0.033 & 0.644 & -0.018 & 0.033 & 0.898 & 0.192 & -0.138 & -0.866 & 0.999  & 0.0006    \\    
Error             & 0.017 & 0.003 & 0.029 & 0.013  & 0.135 & 0.003  & 0.008 & 0.043 & 0.015 & 0.020  & 0.053  & $(\ldots)$& $(\ldots)$   \\       
$m_1,  y - J$     & 0.028 & 0.031 & 0.035 & -0.042 & 0.957 & -0.023 & 0.030 & 1.163 & 0.492 & -0.267 & -1.566 & 1.002  & 0.0003 \\
Error             & 0.017 & 0.003 & 0.020  & 0.010  & 0.167 & 0.003 & 0.007 & 0.033 & 0.039 & 0.022 & 0.123 &  $(\ldots)$ & $(\ldots)$  \\
$[m],  y - J$     & 0.044 & 0.034 & 0.082 & -0.098 & 0.530 & -0.031 & 0.040 & 1.140 & 0.286 & -0.186 & -1.027 & 0.998  & 0.0006     \\
Error             & 0.015 & 0.003 & 0.033 &  0.020 & 0.147 &  0.004 & 0.008 & 0.050 & 0.026 &  0.029 &  0.084 & $(\ldots)$ & $(\ldots)$  \\
$m_1,  y - H$     & 0.041 & 0.029 & 0.008 & -0.019 & 1.071 & -0.017 & 0.028 & 0.932 & 0.330 & -0.179 & -1.326 & 1.002  & 0.0012 \\
Error             & 0.020 & 0.007 & 0.034 &  0.014 & 0.394 &  0.005 & 0.018 & 0.050 & 0.052 &  0.028 &  0.218 & $(\ldots)$ & $(\ldots)$  \\
$[m],  y - H$     & 0.088 & 0.019 & -0.027 & -0.012 & 1.101 & -0.011 & 0.013 & 0.875 & 0.254 & -0.164 & -1.125 & 0.998  & 0.0010     \\
Error             & 0.022 & 0.005 &  0.039 &  0.018 & 0.212 &  0.004 & 0.011 & 0.059 & 0.025 &  0.025 &  0.094 & $(\ldots)$ & $(\ldots)$  \\
$m_1,  y - K$     & 0.029 & 0.035 & 0.031 & -0.028 & 0.735 & -0.021 & 0.040 & 0.876 & 0.245 & -0.139 & -1.066 & 0.999  & 0.0004     \\
Error             & 0.005 & 0.002 & 0.008 &  0.003 & 0.093 &  0.001 & 0.005 & 0.014 & 0.010 &  0.007 &  0.047 & $(\ldots)$ & $(\ldots)$  \\
$[m],  y - K$     & 0.100 & 0.024 & -0.023 & -0.018 & 0.880 & -0.014 & 0.018 & 0.825 & 0.193 & -0.133 & -0.912 & 1.000  & 0.0006    \\
Error             & 0.011 & 0.003 &  0.020 &  0.009 & 0.123 &  0.003 & 0.007 & 0.030 & 0.011 &  0.012 &  0.050 & $(\ldots)$ & $(\ldots)$  \\

\hline                                   
\end{tabular}
\end{table*}

\section{Conclusions}
We have presented a new theoretical metallicity calibration based on the $m_1$
index and on visual--NIR colors to estimate the global metal abundance of cluster and 
field dwarf stars.
We adopted $\alpha$-enhanced evolutionary models transformed into the observational 
plane by using atmosphere models with the same chemical mixture to derive the
new MIC relations. This is the first time that visual--NIR colors are adopted to 
estimate photometric metallicities of dwarf stars. 
The main advantages of the new MIC relations are the following:

{\em i\/}) the molecular bands $CH$, $CN$ and $NH$ affect the $m_1$ index, but
the color indices are unaffected by their presence;

{\em ii\/}) the metallicity sensitivity is larger in the metal-rich regime, 
i.e. for $\mh \gtrsim$ -1.0, where isochrones in the $m_1$ versus 
$y$ - NIR color planes are well separated compared to the same 
isochrones in the $m_1$ versus \strom color planes. The sensitivity 
is larger also in the metal-poor regime, where isochrones overlap 
in the $m_1$ versus \strom color planes at bluer colors, i.e. for $\bmy \le$ 0.4 mag;

{\em iii\/}) the slopes of the MIC relations based on visual--NIR colors are on average 
shallower than the MIC relations based on \strom colors. This means that the former 
indices have, at fixed $m_1$ value, a stronger temperature sensitivity;

{\em iv\/}) they do not include the $u$ filter, which is very time consuming 
from the observational point of view. Moreover, \strom
photometry in the $u$-band is usually less accurate given the reduced sensitivity 
of CCDs in this wavelength region.

In order to validate the new theoretical metallicity calibration 
we adopted two sample of field dwarfs with \strom and NIR photometry and 
high-resolution spectroscopy available. 
The first sample includes 96 dwarfs selected from the study by AR10.
The mean difference between photometric and spectroscopic abundance
is $-0.02\pm0.10$ dex, with a mean intrinsic dispersion of $\sigma$ = 0.31 dex 
($m_1,\ y-J$, $m_1,\ y-H$, $m_1,\ y-K$ relations),
while is $0.07\pm0.06$ dex, with $\sigma$ = 0.31 dex ($[m],\ y-J$, 
$[m],\ y-H$, $[m],\ y-K$ relations).

A further check has been performed by adopting 185 dwarfs selected from
the study by CAS11. In this case the mean difference between photometric 
and spectroscopic abundance is $-0.06\pm0.07$ dex, with 
a mean intrinsic dispersion of $\sigma$ = 0.22 dex ($m_1,\ y-J$, 
$m_1,\ y-H$, $m_1,\ y-K$ relations), while is $-0.01\pm0.06$ dex, 
with $\sigma$ = 0.25 dex($[m],\ y-J$, $[m],\ y-H$, $[m],\ y-K$ relations).

The quoted independent comparisons indicate that the new theoretical 
MIC relations provide accurate metal abundances for field dwarf stars 
with a dispersion smaller than 0.3 dex.

We tested the calibration by adopting also MS stars of two 
GGCs covering a broad range in metal abundance, i.e. 
M~92 ($\feh$ =  - 2.31) and 47~Tuc ($\feh$ = -0.72), 
for which both \strom and WFC3 NIR photometry were available. 

We found that the metallicity distributions of 47~Tuc based on the new visual--NIR 
MIC relations are larger than suggested by spectroscopic 
measurements of both iron and $\alpha$-element abundances. Most of the spread
is given by photometric errors but a fraction of it, and in particular the asymmetry in the metallicity 
distribution might be caused by the occurrence of multiple populations in this 
cluster \citep{milone12}. 
Unfortunately, the stars for which we estimated the metallicity do not have, 
to our knowledge, spectroscopic measurements of iron, $\alpha$ and light elements. 
Therefore, we cannot constrain,  on a quantitative basis, whether the large spread and the
asymmetry in the metallicity distribution of 47~Tuc MS stars is caused by peculiar abundance 
patterns. According to current estimates photometric errors either in optical or in 
NIR magnitudes increase the spread in metallicity by at most for 0.30 dex.  
New medium and high-resolution spectra for the selected cluster MS stars can help 
us to shed new light on the culprit(s) causing the large spread in metal abundance.  

In the case of M~92 the spread of the metallicity distributions is given by the large
photometric errors of the catalog at this magnitude level. 

We also plan to provide independent calibrations of the new visual-NIR diagnostic 
by using either semi-empirical CTRs for both \strom \citep{clem} and NIR 
colors and/or colors predicted by different sets of atmosphere models 
(PHOENIX, Hautschild et al. 1999a,b; MARCS, Gustafsson et al. 2008) 
or different sets of evolutionary models \citep{dotter07,dotter08, girardi, vandenberg}.

\begin{acknowledgements}
      We acknowledge Luca Casagrande for kindly sending us photometric and spectroscopic 
      data for field dwarfs and for the very useful suggestions that helped us to improve 
      the content of the paper.
      We thank Santi Cassisi for his constructive suggestions and advices.
      Support for this work has been provided by the IAC (grant 310394), and the Education and Science 
      Ministry of Spain (grants AYA2007-3E3506, and AYA2010-16717).
      We acknowledge the referee for his/her pertinent comments 
      and suggestions that helped us to improve the content and the 
      readability of the manuscript.
\end{acknowledgements}


\end{document}